\documentclass[article,aps,prd,9pt,notitlepage,twocolumn]{revtex4-2}

\usepackage{blindtext}
\usepackage{graphicx}
\usepackage{amsmath,graphicx}
\usepackage{xcolor}
\usepackage{subcaption}
\usepackage{commath}
\usepackage{float}
\usepackage{fullpage}
\usepackage{appendix}
\usepackage{ragged2e}

\usepackage{bm}
\usepackage{amssymb}
\usepackage{graphicx,xcolor}
\usepackage{csquotes}
\usepackage{amsmath,graphicx}

\begin{document}
\def\la{{\langle}}
\def\u{\hat U}
\def\A{\mathcal A}
\def\PP{\mathcal P}
\def\B{\hat B}
\def\C{\hat C}
\def\D{\hat D}
\def\R{\text{Re}}
\def\Im{\text{Im}}
\def\Om{\Omega}
\def\Omm{{[a,b]}}
\def\up{\uparrow}
\def\do{\downarrow}
\def\fb{\overline F}
\def\pia{\hat \pi_A}
\def\wb{\overline W}
\def\nl{\newline}
\def\m{_{com}}
\def\h{\hat H}
\def\G{\Gamma}
\def\lm{\lambda}
\def\lmu{\underline\lambda}
\def\q{\quad}
\def\t{\tau}
\def\xx{ x^{cl}}
\def\om{\omega}
\def\s{\mathcal{S}}
\def\r{\color{black}}
\def\g{\color{green}}
\def\b{\color{blue}}
\def\n{\\ \nonumber}
\def\ra{{\rangle}}
\def\Ep{{\mathcal{E}}}
\def\E{E_r}
\def\Ee{{\epsilon}}
\def\a{{\hat a}}
\def\sx{{\hat \sigma_x}}
\def\sy{{\hat \sigma_y}}
\def\sz{{\hat \sigma_z}}
\def\h{\hat{H}}
\def\ha{\hat{H}_A}
\def\ua{\hat U_A}
\def\uu{\hat u}
\def\rh{{\rho}}
\def\hh{{\mathcal h}}
\def\e{\enquote}
\def\d{{e}}
\def\a{{\alpha}}
\def\nn{n_1}
\def\nnn{\overline{n}_1}
\def\nm{\overline{n}}

\title{Quantum measurements and delays in scattering by zero-range potentials }
\author{X. Guti\'errez de la Cal$^{a,d}$}
\author{M. Pons$^{b,d}$}
\author{D. Sokolovski$^{a,c,d,*}$}
\affiliation{$^{a}$ Departamento de Qu\'imica-F\'isica, Universidad del Pa\' is Vasco, UPV/EHU,  48940 Leioa, Spain}
\affiliation{$^{b}$ Departamento de F\' isica Aplicada, Universidad del Pa\' is Vasco, UPV-EHU, 48013 Bilbao, Spain}
\affiliation{$^{c}$ IKERBASQUE, Basque Foundation for Science, E-48011 Bilbao, Spain}
\affiliation{$^{d}$ EHU Quantum Center, Universidad del Pa\' is Vasco, UPV/EHU, 48940 Leioa, Spain}
\email{dgsokol15@gmail.com}
 \date\today
\begin{abstract}
{Eisenbud-Wigner-Smith delay and the Larmor time give different estimates for the duration of a quantum scattering event.
The difference is most pronounced in the case where de-Broglie wavelength is large compared to the size of the scatterer.
We use the methods of quantum measurement theory to analyse both approaches, and to decide which one of them, if any,
describes the duration a particle spends in the region which contains the scattering potential.
The cases of transmission, reflection and three-dimensional elastic scattering are discussed in some detail.}
\end{abstract}
\maketitle

\section{Introduction}
It is only natural to expect  a quantum scattering process, be it collision between two particles, or tunnelling across a potential barrier
to be characterised by a particular  duration. Discussion about what this duration should be, and how is ought to be measured, continues to date
\cite{Field}. Traditionally, there were two schools of thought.  An approach, originally due to Eisenbud and Wigner  \cite{Wig1},
and later extended by Smith to multichannel scattering \cite{Smith}, relies on propagation of wave packet states, and leads to time parameters
expressed as energy derivative of the phase of a scattering amplitude. An alternative method, first proposed by Baz' \cite{Baz1}, and later developed in  \cite{BZP},
employed a spin, precessing in a small magnetic field introduced in the region of interest. The Larmor times, obtained in this manner,
involved variations of scattering amplitudes in response to a small constant potential, added in the region \cite{Butt, DSBr}. The authors of \cite{BZP}
concluded that  the approach  \cite{Smith} is in general incorrect, since both methods often lead to similar, yet not identical results \cite{BZP}.
\newline
It is reasonable to ask whether Eisenbad-Wigner-Smith (EWS) approach is merely wrong, or if, perhaps, one deals with two different yet equally valid methods?
Similar questions have been asked, e.g., in \cite{Haug, Land}, and more recently in \cite{Att}.
{\r The reader may also be interested in Refs.\cite{R1}-\cite{R4}}
\newline
{\r The appearance of both Eisenbud-Wigner-Smith and Larmor times may look strange to anyone used to the averages obtained with the help of  a probability distribution, since neither of the two parameters look like conventional averages. 
The problem is most easily understood in terms of quantum measurement theory. In both cases the particle is pre- an post-selected 
in its initial and final (transmitted) states.
In both cases one evaluates, in the standard way, an average of a variable expected to 
contain information about the duration spent in the barrier, a spin's component, or the final particle's position. 
A connection with quantum measurements is established once  one notes that the probabilities used for the averaging 
are given by a convolution of an amplitude distribution with a kind  of "apparatus function".
In the Larmor case, the amplitude distribution refers to the duration of a Feynman's path spent in the barrier region, 
and the apparatus function is determined by the initial state of the clock (see, e.g.,  \cite{DSel}). 
The EWS case is less obvious, but similar. The EWS amplitude distribution describes the range of spacial shifts with which a particle with 
a known momentum may emerge from the barrier, and the apparatus function is the envelope of the initial wave packet state (see, e.g., \cite{DSNT}). 
At this point one notes an important role played by the Uncertainty Principle (UP)  \cite{FeynL}. Since tunnelling can be seen as a result 
of destructive interference between alternatives, the presence of the apparatus function destroys this interference and, 
with it, the studied transition. The only way to preserve the transition is to make the apparatus function very broad, but then the UP
would forbid one to distinguish between the durations, or the shifts involved \cite{FeynL}. This is, indeed the case, since 
if the perturbation is minimised, the measured average is expressed in terms of the first moment  of an {\it amplitude} (and not a {\it probability}) distribution (also known as the "weak value" \cite{WVst,DSw}). In addition to being complex valued, the distribution may change sign, and the "weak value" does not faithfully represents the 
the range of values available to the transition  \cite{DSel}. For example, EWS time, measured in this manner, can turn out to be anomalously short, 
even though the barrier provides only for delays, compared to free propagation \cite{DSNT}.
\newline
In this paper we use the measurement theory techniques  in order to analyse the similarities and the differences between both methods for determining the "tunnelling time" \cite{FOOT}. We will also show that the same approach can be applied to reflected particles, as well as in the case of potential scattering, which was the subject originally studied in  \cite{Wig1}. 
As an illustration, we consider particles scattered by a zero-range potential \cite{ZRP},  chosen for two main reasons.
Firstly, the disagreement between the Larmor and the Eisenbud-Wigner-Smith approaches is is most pronounced in the ultra-quantum case where the particle's de Broglie wavelength exceeds the size of the scatterer. Secondly, in both cases the amplitudes distributions, on 
which our analysis is based, have a particularly simple form,
and the narrative can be abbreviated accordingly.}
\newline
The rest of the paper is organised as follows.
In Sect.II we describe two methods for measuring the duration  $\t$ a classical particle spends in a region containing the scattering potential.
In Sect.III we briefly review the Larmor clock method and show that it predicts a zero delay whenever the size of the scatterer vanishes.
In Sect.IV we show that following the centre of mass of the scattered state leads to a different kind of \e{quantum measurement}.
An inaccurate measurement of this kind determines a Eisenbud-Wigner-Smith time delay, which does not vanish for a zero-range potential.
Sections V and VI  analyse the centre-of-mass delay in transmission across zero-range barrier or well.
In Sect. VII we extend the analysis to reflected particles. In Sect. VIII  we consider elastic scattering by a zero-range spherically symmetric potential. Section IX contains our conclusions.
\section{Two ways to measure a classical duration }
\begin{figure}[H]
\includegraphics[width = \linewidth]{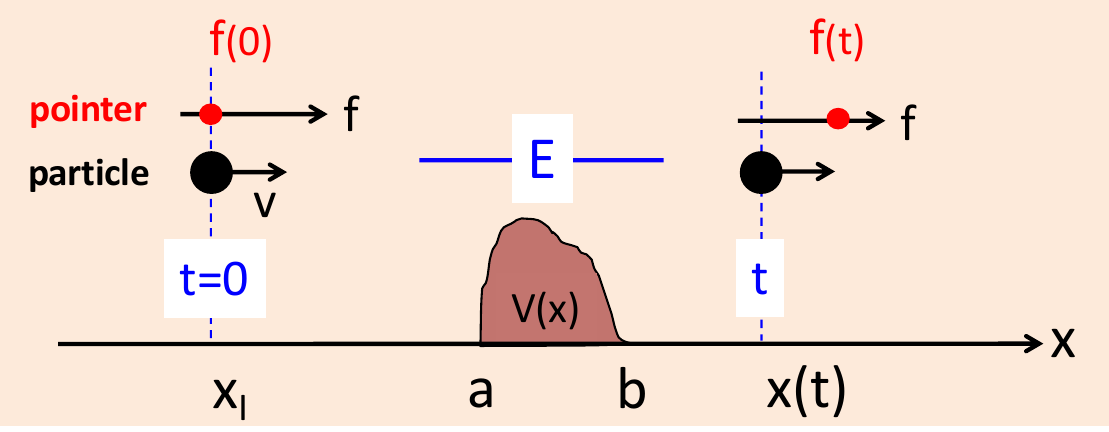}
\centering
\captionof{figure}{A classical particle is coupled to a clock. Eq (\ref{1}), which runs only while it is inside the potential. The duration the particle has spent in $\Omm$ can be read off the pointer's position [cf. Eq.(\ref{2})].
 }
\label{fig_etaTR_1}
\end{figure}

\begin{figure}[H]
\includegraphics[width = \linewidth]{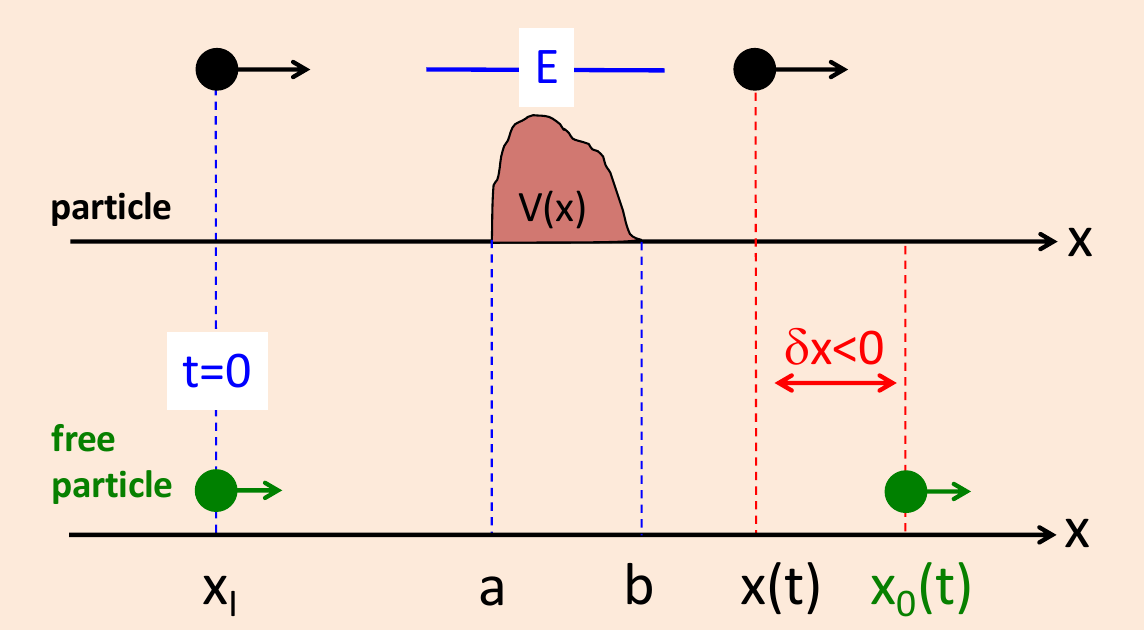}
\centering
\captionof{figure}{The same duration can be evaluated by launching two classical particles, one with and one without the potential,
and comparing their positions once they have crossed the region $\Omm$ [cf. Eq.(\ref{3})].}
\label{fig_etaTR_2}
\end{figure}

In classical mechanics, a particle always moves along a trajectory
$\xx(t)$, and the amount of time $\t$ it spends in
a region $\Omm$,  containing a potential barrier (or a well) $V(x)$ (see Fig. \ref{fig_etaTR_1}),  is a well defined and useful concept. One way to measure it is to couple the particle to a clock, which runs only while the particle is in the region. This can be achieved by equipping the particle with a magnetic moment, introducing a magnetic field in $\Omm$, and dividing the angle of the precession $\phi$ by the Larmor frequency $\om_L$. A simpler version of the Larmor clock is  a pointer with position $f$ and momentum $\lm$, coupled to the particle while it is in the region. The full Hamiltonian of the system, therefore, is
\begin{eqnarray} \label{1}
H(x,p,f,\lm) = p^2/2m+V(x) + \lm \Theta_\Omm (x),
\end{eqnarray}
where $x$ and $p$ are the particle's position and momentum, respectively, and $\Theta_\Omm (x)$ is unity if $x$ lies inside the interval $\Omm$, and $0$ otherwise.
Solving the Hamilton's equations for $\lm=0$ one easily finds the final pointer's reading $f$ equal to the sought duration,
\begin{eqnarray} \label{2}
f(t)-f(0) =\int_0^t\Theta_\Omm (\xx(t'))dt'= \n
\int_a^b m^{1/2}dx/\sqrt{2[E-V(x)]}\equiv \tau(E),
\end{eqnarray}
where $E>V(x)$ is the particle's energy.
The same quantity can be determined without the help of a clock, simply by comparing the current  positions of the particle moving with and without the potential, and define (we reserve the subscript $0$ for free motion)
\begin{eqnarray} \label{2a1}
\delta \xx =\xx(t) - \xx_0(t)
\end{eqnarray}
where $\xx_0(t)$ is the trajectory with $V(x)=0$ (see Fig. \ref{fig_etaTR_2}).
Indeed, if  both particles are launched simultaneously with equal initial momenta $p=\sqrt{2mE}$ from the same initial position, $x_I<a$,
we have $ \xx_0(t)=vt+x_I$, where $v=p/m$.
Both particles cross the region $\Omm$ since  $E>V(x)$.  By the time the particle slowed down by a barrier
reaches $x=b$, its  faster free moving counterpart will lie ahead by $ |\delta x^{cl}|= v(\tau -\tau_0)$, where $\tau_0= (b-a)/v$ is the duration
the free particle spends in the region. The difference in positions can, therefore,  be used to evaluate the $\tau$ defined earlier in Eq.(\ref{2}),
\begin{eqnarray} \label{3}
\tau(E)=  \tau_0(E)-\delta x/v,
\end{eqnarray}
A particle crossing a potential well, $V(x)<0$, spends in $\Omm$ a shorter duration, $\delta x^{cl}>0$, and $\tau(E) < \tau_0(E)$.
Note that since both particles move freely for $x>b$, the distance between them remains the same once the region containing the potential  is crossed.
Note also that this is also a kind of a measurement, where the role of the pointer is now played by the particle itself.
Both approaches can be generalised to quantum scattering, albeit with different results, and we will consider them separately.
\section{The quantum Larmor time}
The quantum analogue of Eq.(\ref{2}) has often been discussed before (see, e.g., \cite{DSL}), and we will describe it here only briefly.
For a quantum particle, the transition amplitude between the states $|\psi\ra$ and $|\phi\ra$ is given by Feynman  path integral (we are using $\hbar=1$)
$A(\phi,\psi,t) =\sum_{paths}\phi^*(x') \exp[iS(x,x',t)] \psi(x)$, where $S$ is the classical action functional, and
$\sum_{paths}=\int dx'\int dx \int_{(x,0)}^{(x',t)}Dx(t)$ includes summation over all virtual paths connecting $(x,0)$ with
$(x',t)$, as well as integration over the initial and final position. The classical expression in  Eq.(\ref{2}) can be promoted to a functional
on the Feynman's paths,
$t_\Omm[x(t)]=\int_0^t \Theta_\Omm(x(t')) dt'$, and
\begin{eqnarray} \label{1a}
A(\phi,\psi,t|\tau) = \sum_{paths}\delta \left (t_\Omm[x(t)]-\tau\right)\times \n
 \phi^*(x') \exp(iS(x,x',t) \psi(x)
\end{eqnarray}
yields the probability amplitude to complete the transition, while spending $\tau$ seconds in the chosen region $\Omm$.
Expressing the Dirac delta $\delta (t_\Omm[x(t)]-\tau)$ as a Fourier integral allows one to rewrite (\ref{1a}) as
\begin{eqnarray} \label{2a}
A(\phi,\psi,t|\tau) = (2\pi)^{-1}\int_{-\infty}^{\infty} d\lm \exp(i\lm \tau)\times \n
 A_{V(x)+\lm \Theta_\Omm(x)}(\phi,\psi,t),
\end{eqnarray}
where $A_{V(x)+\lm \Theta_\Omm(x)}(\phi,\psi,t)$ is the same transition amplitude, but in the modified potential $V(x)+\lm \Theta_\Omm(x)$.
Note that there is a kind of uncertainty relation: to know $\tau$ one needs to make the potential in $\Omm$ uncertain.
In the case of transmission (T) across the barrier [cf. Fig.\ref{fig_etaTR_1}] the transition is between the same positive momentum states,
$|\psi\ra=|\phi\ra =|p\ra$, over a long time, $t\to \infty$, and $A(\phi,\psi,t)$ is just the barrier's transmission amplitude, $T(p,V)$.
Thus, quantum transmission is characterised by a range of durations, each endowed with an amplitude
\begin{eqnarray} \label{3a1}
A_T(p,\tau) = (2\pi)^{-1}\int_{-\infty}^{\infty} d\lm \exp(i\lm \tau)\times \n
 T(p,V+\lm \Theta_\Omm).
\end{eqnarray}
In its quantum version, the clock  (\ref{1}) may be prepared in an initial state $|G\ra$, e.g., a Gaussian $G(f)= \la f|G\ra= C \exp(-f^2/\Delta f^2)$, centred at $f=0$.
The pointer would be displaced by $\tau$, $G(f) \to G(f-\t)$, if the value of $\tau$ were unique.
With many such values, the final state of the clock
is given by a sum over all displacements,
\begin{eqnarray} \label{3a}
\Phi(f) = \int_0^\infty G(f-\tau)A_T(p,\tau)d\tau.
\end{eqnarray}
{\r 

* Equation (\ref{3a}) defines the measurement briefly discussed in the Introduction.

* Equation (\ref{2}) defines $\tau$ as the {\it net duration} spent by the particle in the barrier region.

* Since $T(p,V)=\int A_T(p,\tau)d\t$,  different durations interfere and cannot be told apart without a clock.

* The amplitude of finding the particle in a final state $|p\ra$, and the clock's pointer in $|f\ra$ is the 
same as that of finding the particle in $|p\ra$ provided the durations spent in the barrier 
were restricted to a range  $f-\Delta f \lesssim \tau \lesssim f+\Delta f$.
One can say that $\tau$ has been measured to an accuracy $\Delta f$}.

The clock is the more accurate the smaller is $\Delta f$. It is also more perturbing, and sending $\Delta f \to 0$ would quench transmission, causing the  particle to be reflected.
In the opposite limit $\Delta f\to \infty$, {\r an individual clock reading $f$ provides little information. 
However, using (\ref{3a}) one can calculate the mean pointer reading (see Appendix A)
\begin{eqnarray} \label{3ac}
\la f \ra \equiv \frac{ \int f |\Phi(f)|^2  df}{ \int |\Phi(f)|^2  df}{\xrightarrow[ \Delta f \to \infty ] {}} \R [\overline{\t}_{\Omm}(p)].
\end{eqnarray}
where $\overline{\t}_{\Omm}(p)$ is the "complex
 time" of Ref.\cite{DSB},
\begin{eqnarray} \label{3ab}
\overline{\t}_{\Omm}(p) =\frac{ \int_0^\infty \tau A_T(p,\tau) d\t}{\int_0^\infty A_T(p,\tau) d\t} = \n
-i \partial_\lm\left [ \ln  T(p,V+\lm \Theta_\Omm)\right ]|_{\lm=0}.
\end{eqnarray}
Evaluated with an alternating complex valued distribution, the \e{weak value} \cite{WVst} $\overline{\t}_{\Omm}(p)$  does not have the properties of a physical time interval,
in agreement with the Uncertainty Principle \cite{FeynL} which forbids knowing the duration $\t$ in precisely the 
same sense it forbids knowing the slit chosen by the particle in a double-slit experiment  \cite{DSel}. 
}
\newline
Our main interest here is in the delay (if any), experienced by a particle scattered by a zero-range potential,
\begin{eqnarray} \label{4a}
V(x)=U \Theta_\Omm (x), \q (b-a) \to 0, \n
\q U \to \infty, \q U(b-a)=\Om =const.
\end{eqnarray}
As the region becomes ever more narrow, $a\to b$, $t_\Omm[x(t)]$ can only tend to zero for any smooth path $x(t)$.
However, Feynman's paths are notoriously irregular \cite{FeynH}, and a more rigorous justification will be given next.
If the amplitude distribution for a free particle, $V(x)=0$, $A_0(p,\tau)$ is known, the distribution for a rectangular
potential $V(x)=U \Theta_\Omm (x)$, barrier or well,  takes a particularly simple form,
\begin{eqnarray} \label{5a}
 A_T(p,\tau) = \exp(-iU\tau) A_0(p,\tau).
\end{eqnarray}
$A_0(p,\tau)$ can be computed by closing the integration contour
in Eq.(\ref{3a1}) (with $V=0$)  in the upper half of the complex $\lm$-plane, where $T(p,\lm \Theta_\Omm)$ has poles \cite{DSBr}.
Only one pole contribution survives in the limit  $(b-a) \to 0$, and using (\ref{5a}) one finds \cite{DSBr}
\begin{eqnarray} \label{6a}
A_T(p,\tau) {\xrightarrow[ a\to b ] {}}\t_0^{-1} \exp[-i \Om\tau/(b-a)]\times \n
\exp(-\t/\t_0),
\end{eqnarray}
where $ \t_0=m(b-a)/p =(b-a)/v$. Since   $|A_T(p,\tau)|$ tends to $\delta(\tau)$, a measurement  by a Larmor clock
will always yield a zero duration for a very narrow potential [cf. Eqs. (\ref{3a1})-(\ref{3ab})].
Next we ask whether the same is true if one tries to deduce the same duration from the final position of a transmitted particle.
\section{The Eisenbud-Wigner-Smith (phase) time}
To obtain a quantum analogue of the procedure, described by Eq.(\ref{3}), one can replace  classical particles by wave packets (WP),
and evaluate the distance between their centres of masses (COMs)  as shown in Fig.3.
\begin{figure}[H]
\includegraphics[width = \linewidth]{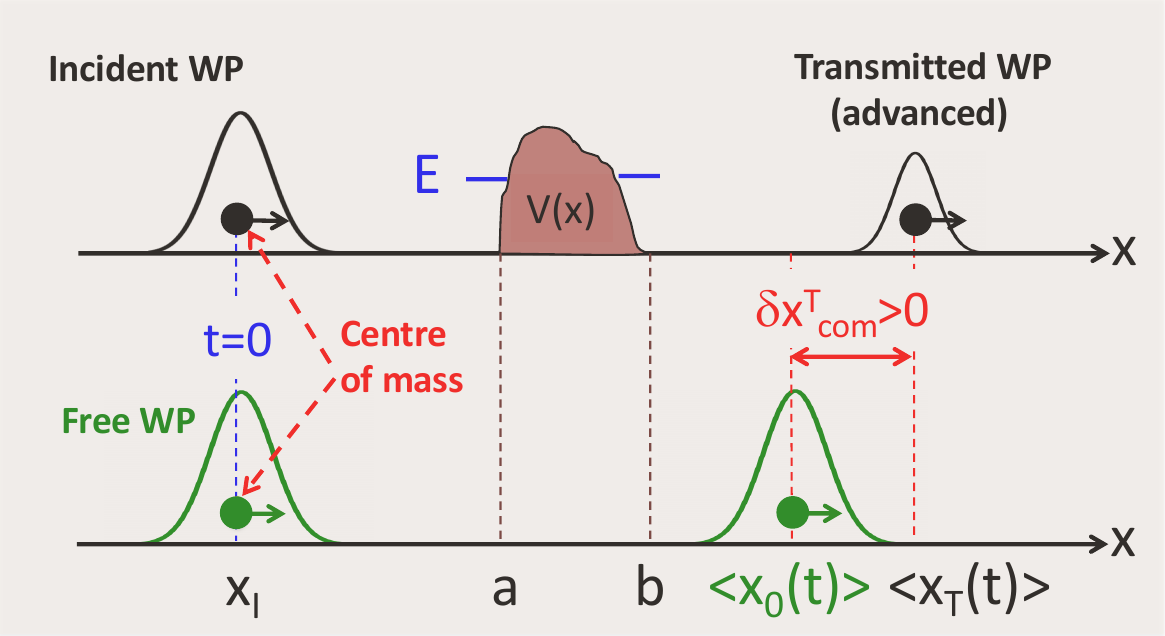}
\centering
\captionof{figure}{ A quantum analogue of Fig.2: classical particles are replaced by wave packets, whose centres of mass
are used as the reference points. The advancement, or delay, of the transmitted particle results from the interference between
all virtual spatial shifts provided by the potential [cf. Eq.(\ref{3b})].
 }
\label{fig_etaTR_3}
\end{figure}
At $t=0$, both particles are prepared in the same state with a mean momentum $p$, a width $\Delta x$, and the COM at some $x_I<0$,
$|x_I |>>\Delta x$.
After scattering, the transmitted and the freely propagating state both lie to the right of the barrier.  They  are given by
\begin{eqnarray} \label{-1b}
\psi_T(x,t) =\int T(k,V)A(k,p)\times \n
\exp(ikx -iE_kt)dk, \q E_k=k^2/2m,
\end{eqnarray}
and
\begin{eqnarray} \label{0b}
\psi_0(x,t) =\int A(k,p) \exp(ikx -iE_kt)dk= \n
\exp(ipx-iE_pt)G_0(x,t),
\end{eqnarray}
where $G_0(x,t)$ is an envelope of a width $\Delta x$,, initially peaked around $x=x_I$.
For the separation between their COMs, we have
\begin{eqnarray} \label{1b}
\delta x^T\m \equiv \la x(t)\ra_T -\la x(t)\ra_0,\n
\la x(t)\ra_{T,0} \equiv \frac{\int x|\psi_{T,0}|^2 dx}{\int |\psi_{T,0}|^2 dx},
\end{eqnarray}
where, by Ehrenfest's theorem \cite{Ehre},
\begin{eqnarray} \label{2b}
 \la x(t)\ra_0= pt/m +x_I\equiv vt+x_I.
\end{eqnarray}
Throughout the paper we will consider what the authors of \cite{Haug} called "completed events", i.e., the situation
where both wave packets move freely to the right of the barrier, and the integration limits in Eq.(\ref{1b})
can be extended to $\pm\infty$.
There is no simple way to use Feynman path integral, as it was done in Eq.(\ref{1a}). However, expressions similar to Eqs.(\ref{2a})-(\ref{3ab})
are readily obtained by rewriting Eq.(\ref{-1b}) as a convolution
\begin{eqnarray} \label{3b}
\psi_T(x,t) = e^{ipx-iE_pt}\times \n
 \int_{-\infty}^\infty G_0(x-x^\prime ,t)\eta_T(x^\prime,p) dx^\prime,
\end{eqnarray}
\begin{eqnarray} \label{4b}
\eta_T(x^\prime,p) = \frac{e^{-ipx^\prime}}{2 \pi} \int_{-\infty}^\infty T(k,V)e^{ikx^\prime} dk,
\end{eqnarray}
which conveniently separates the information of free motion, contained in $G_0$, from the properties  of the scattering potential
which determine $\eta_T(x^\prime,p)$.
{\r If the spreading of the wave packet can be neglected, $G_0(x-x^\prime ,t)\approx G_0(x-x^\prime ,t=0)$ and 
Eq. (\ref{3}) defines the measurement different from the one described by Eq.(\ref{3a}).

* Noting that $\exp(ipx)\eta_T(x^\prime,p)\sim \exp[ip(x-x')]$ allows one to {\it define} $x'$ as 
the spatial shift with which a particle with momentum $p$ emerges from the barrier
(whatever this might mean).

* Since  $T(p,V)=\int \eta_T(x^\prime,p)dx'$ different shifts interfere and cannot be told apart {\it a priori}.

* However, the amplitude of finding a particle, prepared at $t=0$  in a wave packet state (\ref{0b}), at a location $x$
is the same as that of finding a particle, prepared in a state $|p\ra$, in the same state $|p\ra$, 
provided the shifts imposed by the barrier were restricted to a range  $x-\Delta x \lesssim x' \lesssim x+\Delta x$.
Replacing the plane wave $|p\ra$ with a wave packet (\ref{0b}) of a width $\Delta x$ and a mean momentum $p$
allows one to measure  $x'$ to accuracy $\Delta x$.
\newline 
If the measurement is accurate, $\Delta f \to 0$, one recovers the result for the free motion, 
$\psi_T(x,t)\approx \psi_0(x,t)$, since the high momenta which dominate the transmission 
are unaffected by the presence of the scattering potential. Tunnelling, as one may expect, 
is thereby destroyed. 
In the classical limit $\hbar \to 0$, $E(p) > V(x)$, the rapidly oscillating $\eta_T(x^\prime,p)$ develops a stationary
region around $x' =\delta \xx =\xx(t) - \xx_0(t)$ [cf. Eq.(\ref{2a1})] and the classical result is recovered \cite{DSX}. 

The benefits of converting a transmission problem into a quantum measurement one are most evident when discussing the properties of the 
so-called phase time (see, e.g., \cite{Muga}). 
If the $G_0(x,0)$ is very broad (the spreading can be neglected), interference between different shifts is not destroyed, and the Uncertainty Principle allows one
to determine only a \e{complex shift}, $\overline {x'}$, similar to the \e{complex time}, Eq. (\ref{3ab}). Like $\overline{\t}_{\Omm}(p)$ in Eq.(\ref{3ab}), it is obtained by averaging $x'$ with an alternating complex valued distribution (\ref{4b}),
\begin{eqnarray} \label{5b}
\overline {x'}_T(p)= \frac{ \int_{-\infty}^\infty x' \eta_T(x',p) dx'}{\int_{-\infty}^\infty \eta_T(x',p) dx'}\n
= i \partial_p\left [ \ln  T(p,V)\right ].
\end{eqnarray}
Using Eq.(\ref{1b}) one finds (see Appendix A)
\begin{eqnarray} \label{6b}
\delta x^T\m \q {\xrightarrow[ \Delta x \to \infty ] {}}\q \R\left [\overline{x'}_T(p)\right ].
\end{eqnarray}
where $\R\left [\overline{x'}_T(p)\right ]$ does not even have to lie in the region where $\eta_T(x',p)\ne 0$.
One may be tempted to convert the spatial delays into temporal ones using $\delta \t =-x'/v$.
This is justified in a classically allowed case [cf. Eq.(\ref{2a1})] but is unwarranted in general. 
Replacing $\delta x$ in the classical Eq.(\ref{3}) by its quantum analogue (\ref{1b})  yields a \e{phase time} estimate for the duration spent in the barrier region ,
\begin{eqnarray} \label{7b}
\tau^{phase}(p)=(b-a)/v-\R\left [\overline{x'}_T(p)\right ]/v=\n
\frac{b-a}{v}+\frac{1}{v} \frac{\partial \varphi_T(p,V)}{\partial p},
\end{eqnarray}
where $\varphi(p,V)$ is the phase of the transition amplitude, $T(p,V)=|T(p,V)|\exp[i\varphi_T(p,V)]$,
and $v^{-1}\partial \varphi_T(p,V)/\partial p=\partial \varphi_T(p,V)/\partial E$ is the Eisenbud-Wigner-Smith time delay.
\newline
The phase time (\ref{7b}), related to the \e{weak value} of the spacial shift $x'$ in Eq.(\ref{5b}) has the same problem 
as its Larmor counterpart (\ref{3ac}). It can be anomalously short in tunnelling, even though the barrier only delays the particle relative to free propagation
\cite{DSX}. It does not grow as expected if  the barrier width is increased \cite{Hart}. 
 \newline
It is, however, different in one important aspect. 
While the Larmor time vanishes for a zero-range potential, $\tau^{phase}(p)$ remains finite even as $(b-a)\to 0$.
Next we study how an why this happens}

\section{ Gaussian wave packets in a zero-range potential}
The transmission amplitude for a zero-range potential $V(x)=\Om\delta(x)$ (\ref{4a}) is well known to be (below  we put the mass $m$ to unity)
\begin{eqnarray} \label{1c}
T(k,\Om) = 1-\frac{i\Om}{k+i\Om}.
\end{eqnarray}
For $	\Om <0$ its single pole in the complex plane of the momentum lies on positive imaginary $k$-axis, where it corresponds to a bound state.
For $\Om >0$ the pole moves to the negative $k$-axis, and closing in Eq. (\ref{4b}) the contour of integration in the upper or lower half-plane, we obtain 
\begin{eqnarray}\label{2c}
{\eta}_T(x^\prime,p)= 
     \begin{cases}
     \delta(x^\prime) - \Theta(-x') |\Omega| \exp(-ip x^\prime +|\Omega|  x^\prime), \\ \text{if}\q \Om >0\\
     \delta(x^\prime) - \Theta(x') |\Omega| \exp(-ip x^\prime -|\Omega|  x^\prime),  \\ \text{if}\q \Om <0,\\
     \end{cases}
\end{eqnarray}
where $\Theta(x)\equiv 1$ for $x>0$, and $0$ otherwise.
\newline
Equations (\ref{2c}) provide a useful insight into how a potential acts on the incident particle. An incoming plane wave is multiplied by
the transmission amplitude $T(p,\Om)$ and, for a barrier, $\Om > 0$, we have
\begin{eqnarray} \label{3c}
\exp(ipx)\to T(p,\Om)\exp(ipx) =\n
\q\q\q\q\q  \exp(ipx)- \Omega \int_{-\infty}^0 \exp(\Om x') \times \n
 \exp[ip(x-x')]dx'.
\end{eqnarray}
Instead of providing a  {\it temporal} delay for the transmitted particle
(it could do so, e.g., by changing $\exp(ipx-iE_pt)$ into $\exp[ipx-iE_p(t-\tau_p)]$), a barrier acts as an \e{interferometer},
which splits the incoming plane wave into components with different phase shifts, corresponding to  possible {\it spatial} delays,  $x'<0$.
These delays are still present when the width of the barrier goes to zero, provided the strength of the potential increases accordingly
[cf. Eq.(\ref{4a})].
For a well, a similar expression contains additional plane waves, spatially {\it advanced} relative to free propagation.
One classical feature survives in this purely quantum case. In some sense, a barrier tends to \e{delay} the particle, where a well tends to
\e{speed it up}.
\newline
To quantify these effects one can look  at the motion of wave packets.  As always, it is convenient to consider Gaussian states with a mean momentum $p$ and a coordinate width $\Delta x$,
($\Delta k=2/\Delta x$)
\begin{eqnarray} \label{4c}
A(k,p) = 2^{-1/4}\pi^{-3/4}\Delta  k^{-1/2}\times\q\q\q\n
\exp\left[ -(k-p)^2/\Delta k^2-i(k-p)x_I \right],\n
G_0(x,t)=[2\Delta x^2/\pi \sigma_t^4]^{1/4}\exp[-(x-vt-x_I)^2/\sigma_t^2], \n
\sigma_t \equiv(\Delta x^2+2it/m)^{1/2},\n
|G_0(x,t)|=[2/\pi \Delta x_t^2]^{1/4}\exp[ -(x-vt-x_I)^2/\Delta x_t^2],\n
 \Delta x_t\equiv (\Delta x^2+\Delta k^2 t^2/m^2)^{1/2}.
\end{eqnarray}
The analysis is even simpler in the dispersionless case, $E_k=ck$, where the free amplitude undergoes no spreading,
\begin{eqnarray} \label{5c}
\tilde G_0(x,t)=[2/\pi \Delta x^2]^{1/4}\n
\times \exp[-(x-ct-x_I)^2/ \Delta x^2].
\end{eqnarray}
and the similarity between  Eqs.(\ref{3b}) and (\ref{3a}) is yet more evident.
The Larmor clock's pointer state is displaced as a whole without spreading [cf. Eq.(\ref{3a})] because the kinetic energy, $\lm^2/2\mu$ is omitted both
in the classical Hamiltonian (\ref{1}) and in its quantum counterpart (usually, by assuming the pointer's mass $\mu$ to be  large).
A WP with no kinetic energy would not propagate at all, but making $E_k$ linear, rather than quadratic in $k$ has a similar effect.

\section{Centre-of-mass delay for  transmission}
Consider again Eq.(\ref{1b}). According the Heisenberg's Uncertainty Principle, a particle can have either a well defined position,
or a well defined momentum. Therefore,  much depends on the coordinate width $\Delta x$ of the initial Gaussian WP, as well on the dispersion law.
\newline
The dispersionless case (\ref{3b}) is simpler, since the envelope (\ref{5c}), however narrow, is displaced without distortion (see Appendix B).
As $\Delta x \to 0$, only the $\delta$-term in Eq.(\ref{2c}) needs to be taken into account,
$\psi_T(x,t)\approx \psi_0(x,t)$, and  $\delta x\m \to 0$. The contribution from the smooth part of $\eta_T$  vanishes, because
$\int |\tilde G_0(x)|^2dx=1$ for any $\Delta x$, and  $ \int \tilde G_0(x)dx\sim (2\pi \Delta x^2)^{1/4}\to 0$. (A similar situation occurs in quantum measurements,
where a singular $\delta$-term in an amplitude distribution is responsible for Zeno effect \cite{DSerg}.)
This is an expected result, since for $|k|\to \infty$,  $T(k,V)\to 1$, and the potential has no effect on most of the momenta contained in the initial
wave packet.
In the opposite limit, $\Delta x\to \infty$, $\Delta k \to 0$ the singular term in Eq.(\ref{2c}) can be neglected, and  the \e{complex shift} in (\ref{5b})
is determined only by the smooth part of
${\eta}_T(x^\prime,p)$,
\begin{eqnarray} \label{1d}
\overline {x'}_T(p)=  \frac{-\Omega}{p^2 + \Omega^2} + i\frac{\Omega^2}{p(p^2+\Omega^2)} \equiv\n
 \R\left [\overline {x'}_{T}\right ] + i\Im\left [\overline {x'}_{T}\right].
\end{eqnarray}
Its real part, ${-\Omega}/{(p^2 + \Omega^2)}$, yields the distance between the COMs of the two WPs.
The imaginary part of $\overline {x'}_T(p)$ is related to so the called \e{momentum filtering} effect, whereby the mean momentum of the transmitted
WP is increased because  higher momenta are transmitted more easily. Its significance is best illustrated in the case where dispersion is present,  as we will discuss next.
\newline
For $E_k=k^2/2m$ we can consider scattering \e{completed} when the broadened transmitted WP lies sufficiently far to the right of the potential.
The time needed for this can be estimated by following the motion of a free wave packet. We want the initial Gaussian WP placed
as close as possible to the potential, e.g., at  $x_I=-K\Delta x$, $K>1$.
We also want to measure the COMs as soon as the scattering is completed, e.g., when the COM of the freely propagating WP
lies several widths away on the other side of the barrier, e.g.,  at $x_F=K \Delta x_t$, where $ \Delta x_t$ is defined in the last of Eqs.(\ref{4c}).
For a mean momentum $p$ the required time is easily found to be $t(p,\Delta k, K)=2m p\Delta x K/(p^2-K^2\Delta k^2)$.
(Note that we cannot simply send $\Delta x \to 0$, $\Delta k \to \infty$, as was possible for $E_k=ck$).
In the limit $\Delta x \to \infty$ one finds (see Appendix C)
\begin{eqnarray} \label{2d}
\delta x^T_{com}\approx
\R\left [\overline{x'}_T(p)\right ]+\n
 \frac{\Im\left [\overline {x'}_{T}(k)\right]\Delta k^2}{2m}t(p,\Delta k, K)
\end{eqnarray}
The last fraction is clearly the excess mean velocity obtained through the momentum filtering, absent for $E_k=ck$ where all momenta propagate with the same velocity $c$. The last term in Eq.(\ref{2d}) behaves as  $\sim \Delta k\sim 1/\Delta x$, and was omitted in Eq.(\ref{6b}).
It may need to be retained for not-too-broad WPs, as is shown in Fig.4 for $K=3$.
\begin{figure}[H]
\includegraphics[width = \linewidth]{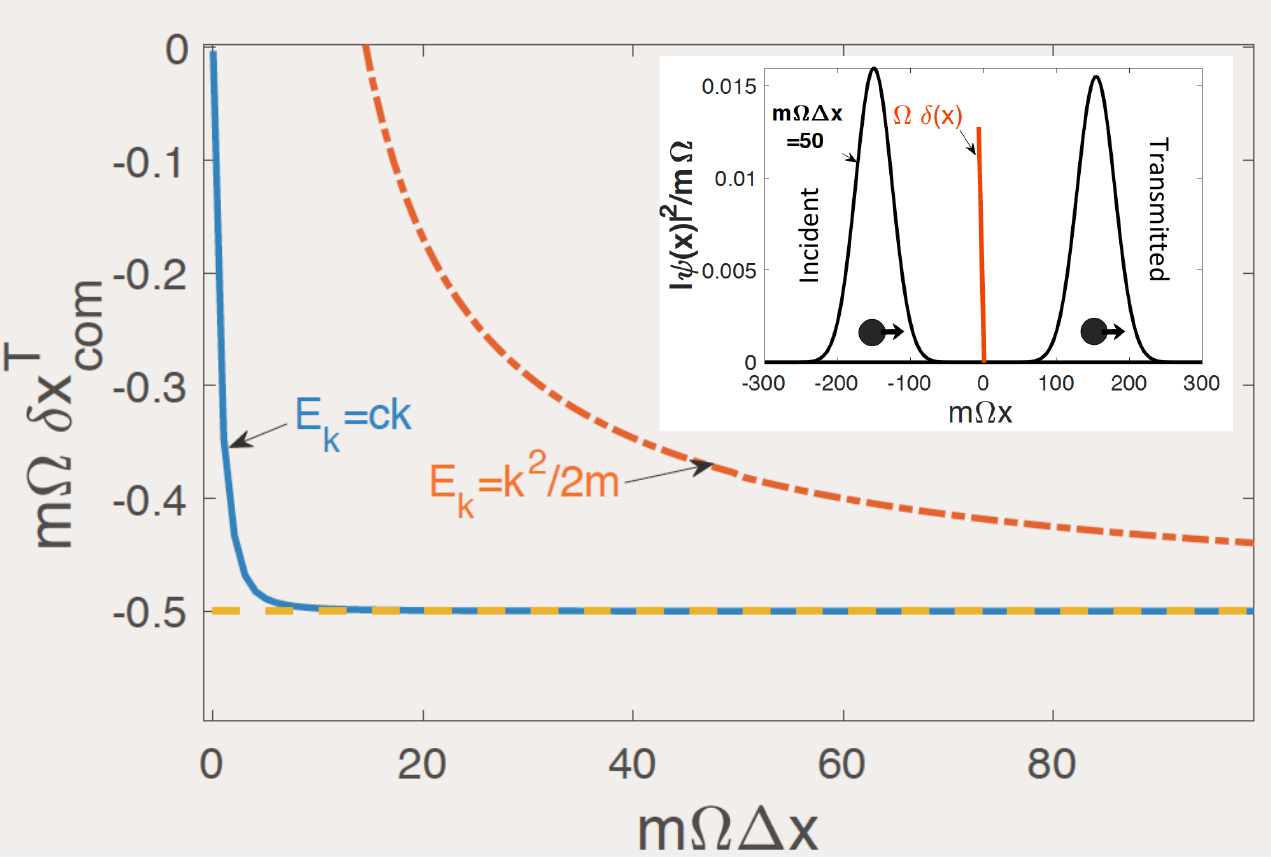}
\centering
\captionof{figure}{ Centre-of-mass delay for transmission by a zero-range barrier, $\Om >0$,
with and without dispersion
vs.  WP's width $\Delta x$.
Shown by the dashed line is the delay in the limit $\Delta x \to \infty$
 [cf. Eq.(\ref{2d})]. The inset shows the initial and final wave packets for $m\Om\Delta x =50$ and $E_k=k^2/2m$.
 The parameters used are: $\Om/c=mc/p=1$, $x_I/\Delta x=3$, and $m\Om^2t(p,\Delta k,3) =6m\Om \Delta x
 /[1-36(m\Om \Delta x)^{
 2}]$.
 }
\label{fig_etaTR_6}
\end{figure}
 Finally, we find the COM of the transmitted WP delayed by a zero-range barrier ($ \delta x_T \approx  \R\left [\overline {x'}_{T}\right] < 0$ if $\Om >0$),
and advanced by a  zero-range well ($\delta x_T \approx \R\left[\overline {x'}_{T}\right]> 0$ if $\Om <0$).
This is not what happens in general, e.g., for a rectangular \cite{DSNT}  or  an Eckart barrier \cite{DSX},
where the COM of the greatly reduced transmitted WP is actually advanced. The reason for this is that in all such cases
$T(k,V)$ has a large (infinite) number of poles in the complex $k$-plane, there are many exponential terms in the r.h.s.
of Eqs.(\ref{2c}), and the resulting ${\eta}_T(x^\prime,p)$ has a complicated form \cite{DSX}. Although it vanishes for $x' >0$,
Gaussian envelopes in Eq.(\ref{1b}) may interfere constructively in a small region of $x > x_0+vt$, and cancel each other elsewhere.
This behaviour  cannot, of course, be reproduced in the much simpler case studied here.
\section{Centre-of-mass delay for reflection}
A similar analysis can be applied in the case of a  particle, reflected (R) by a potential $V(x)$, contained between $x=a$ and $x=b$. The reflected WP is given by
\begin{eqnarray} \label{0e}
\psi_R(x,t) =\int R(k,V)A(k,p)\times \n
 \exp(-ikx -iE_kt)dk,
\end{eqnarray}
where $R(k,V)$ is the reflection amplitude, satisfying $|T(p,V)|^2+|R(p,V)|^2=1$.
It can also be written in a form similar to (\ref{3b})
\begin{eqnarray} \label{1e}
\psi_R(x,t) = e^{-ipx-iE_pt}\times \n
\int_{-\infty}^\infty G_0(-x-x^\prime,t)\eta_R(x^\prime,p) dx^\prime,
\end{eqnarray}
where
\begin{eqnarray} \label{2e}
\eta_R(x^\prime,p) = \frac{e^{-ipx^\prime}}{2 \pi} \int_{-\infty}^\infty R(k,V)e^{ikx^\prime} dk,
\end{eqnarray}
and $ G_0(-x-x^\prime,t)$ (note $x\to -x$) is the envelope of the {
mirror image of the free WP} with respect to the origin,
which is the same (except for a minus sign) as the envelope of a WP reflected by an infinite potential wall at $x=0$.
One can still compare positions of the centres of mass, with and without the potential, by defining
\begin{eqnarray} \label{3e}
\delta x^R\m  =\la x(t)\ra_R +vt+x_I ,\n
\la x(t)\ra_{R} \equiv \frac{\int x|\psi_{R}|^2 dx}{\int |\psi_{R}|^2 dx},
\end{eqnarray}
There is one complication not encountered in the case of transmission. Consider a potential $V_s(x)=
V(x-s)$, obtained by displacing the original barrier or well by a distance $s$. Such a displacement has no
effect on the transmission amplitude, $T(p,V_s)=T(p,V)$, but the reflection amplitude acquires an extra phase,
$R(k,V_s)\to \exp(2iks)R(k,V_s)$, and $\eta_R(x^\prime,p)$ changes into  $\eta_R(x^\prime-2s,p)$.
In other words, one needs to decide where to put the potential before making the comparison with free propagation.
 The ambiguity can be resolved by
always placing the left edge of the potential at the origin, $a=0$ (see Fig.5), and considering the reflected particle {\it delayed}
by the potential if $\delta x^R\m >0$, or {\it advanced} by it, if $\delta x^R\m <0$.

\begin{figure}[H]
\includegraphics[width = \linewidth]{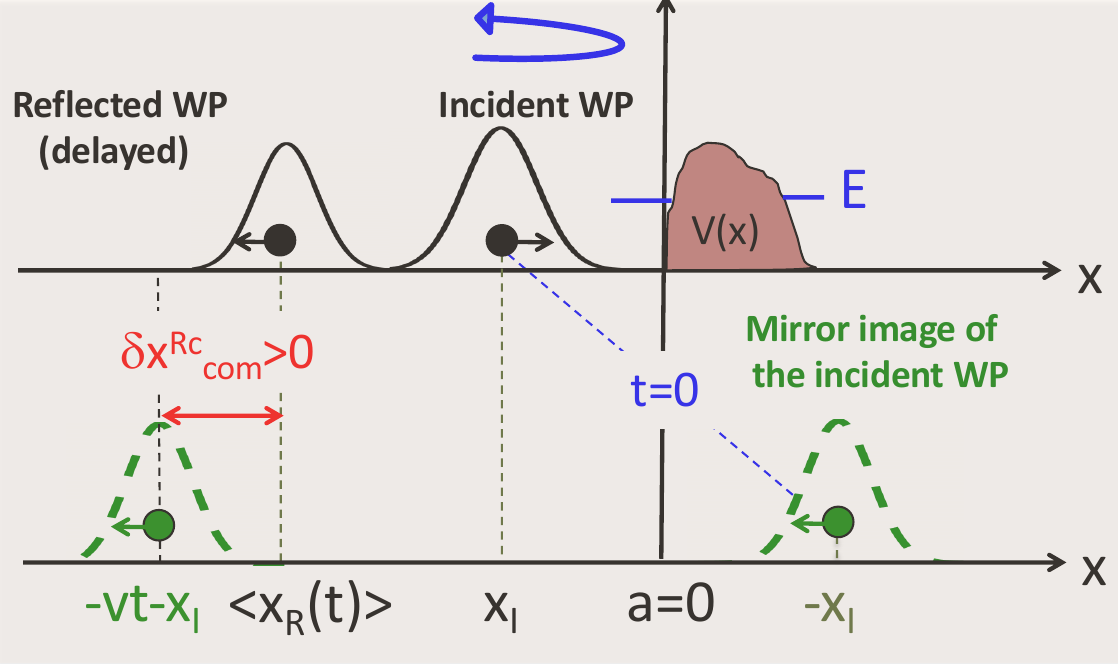}
\centering
\captionof{figure}{The position of a quantum particle, reflected by a potential $V(x)$,
is compared with that of a free particle launched in the opposite direction from $-x_I>0$.
The particle is said to be {\it delayed} by the potential if its COM lies to the right of the COM of the freely
propagating WP, or {\it advanced} if the opposite is true
 [cf. Eq.(\ref{3e})].
 }
\label{fig_etaTR_4}
\end{figure}
\noindent
With this agreed, a classical
reflected particle can only be delayed, since it either bounces off the
edge of the potential at $x=0$, or has to travel further to the right before making a U-turn.
 In the quantum case it is not always the case, as we will show next.
\newline
Reflection amplitude of a zero-range potential $V(x,\Om)=\Om \delta(x)$  is given by
\begin{eqnarray} \label{4e}
 R(k,\Om)=\frac{-i\Omega}{ k + i\Omega},
\end{eqnarray}
and
\begin{eqnarray}\label{5e}
{\eta}_R(x^\prime,p)=
     \begin{cases}
      -\Theta(-x') |\Omega| \exp(-ip x^\prime +|\Omega|  x^\prime), \\ \text{if} \q \Om >0 \\
-\Theta(x') |\Omega| \exp(-ip x^\prime -|\Omega|  x^\prime), \\ \text{if}\q \Om <0.\\ 
     \end{cases}
\end{eqnarray}
[Note the absence of a $\delta$-term, since $R(k,\Om)\to0$ as $|k| \to \infty$.]
According to our convention, a reflected particle with a momentum $p$ would be delayed by a zero-range barrier
(as it would be in a classical case), and advanced by a zero-range well (a purely quantum effect, since there is no reflection from a well
in the classical limit).

As was shown in the previous Section, without spreading [cf. Eq.(\ref{5c})], a narrow wave packet crosses a barrier or a well almost without reflection.
Inserting (\ref{5c}) and (\ref{5e}) into Eq.(\ref{1e}) for the small reflected part we find (the upper sign is for a barrier)
$|\psi_R(x,t)|^2 \sim  \Theta(\pm x\mp ct\mp x_I)\Delta x \exp[-2|\Om|(\pm x\mp ct\mp x_I)]$, and
\begin{eqnarray} \label{5ea}
\delta x^R\m{\xrightarrow[ \Delta x \to 0 ] {}}1/2\Om,
\end{eqnarray}
as shown in Fig. 6.
\begin{figure}[H]
\includegraphics[width = \linewidth]{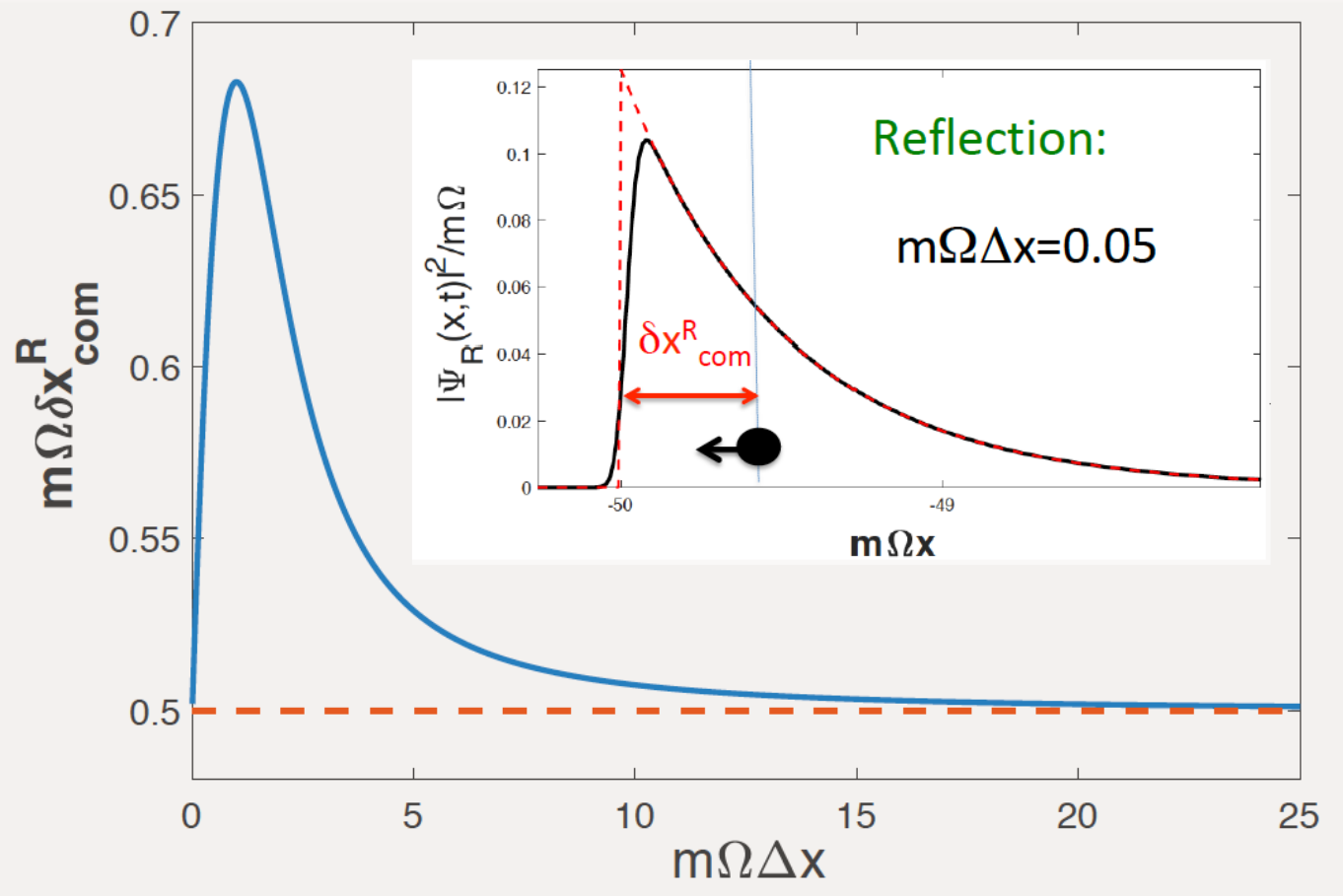}
\centering
\captionof{figure}{Centre-of-mass delay for reflection by a zero-range barrier, $\Om >0$,
in the absence of dispersion, $E_k=ck$, [cf. Eq.(\ref{5c})]
vs.  WP's width $\Delta x$.  Shown by the dashed line is the delay in the limit $\Delta x \to \infty$
 [cf. Eq.(\ref{6e})]. The inset shows the small reflected WP (solid) and its limiting form for $\Delta x \to 0$ (dashed).
 The parameters used are: $\Om/c=mc/p=1$, $m\Om x_I=-50$, and $m\Om ct =100$. }
\label{fig_etaTR_7}
\end{figure}
\noindent
In the opposite limit of a broad WP
we find
\begin{eqnarray} \label{6e}
\delta x^R\m\q {\xrightarrow[ \Delta x \to \infty ] {}}\q \R \left [\overline{x'}_R(p)\right ]= \frac{\Omega}{p^2 + \Omega^2},
\end{eqnarray}
which is valid both with and without dispersion [cf. Eqs.(\ref{4c}) and (\ref{5c})].
In all cases the reflected particle  is delayed if $\Om>0$, and advanced if $\Om <0$.
\section{Centre-of-mass delay in elastic scattering}
Before concluding we revisit the case of a particle scattered by a short-range spherically symmetric potential
$V(r)$ contained between $r=0$ and $r=b$. (One can also think of a collision between two particles interacting
via $V(r)$, $r=|\vec r_1-\vec r_2|$). For a zero angular momentum, $L=0$, one can prepare a specially symmetric
wave packet $\psi(r,t=0)=\int A(k)\exp[-ik(r-r_I)]dk=\exp(-ipr) G_0(r,t=0)$, which converges on the scattering potential.
The state $\psi(r,t)$ satisfies a radial Schr\"odinger equation with a boundary condition $\psi(r=0,t)=0$, and one has
a previously studied case of reflection, with an additional infinite wall added at $r=0$ (see Fig.7).
\begin{figure}[H]
\includegraphics[width = \linewidth]{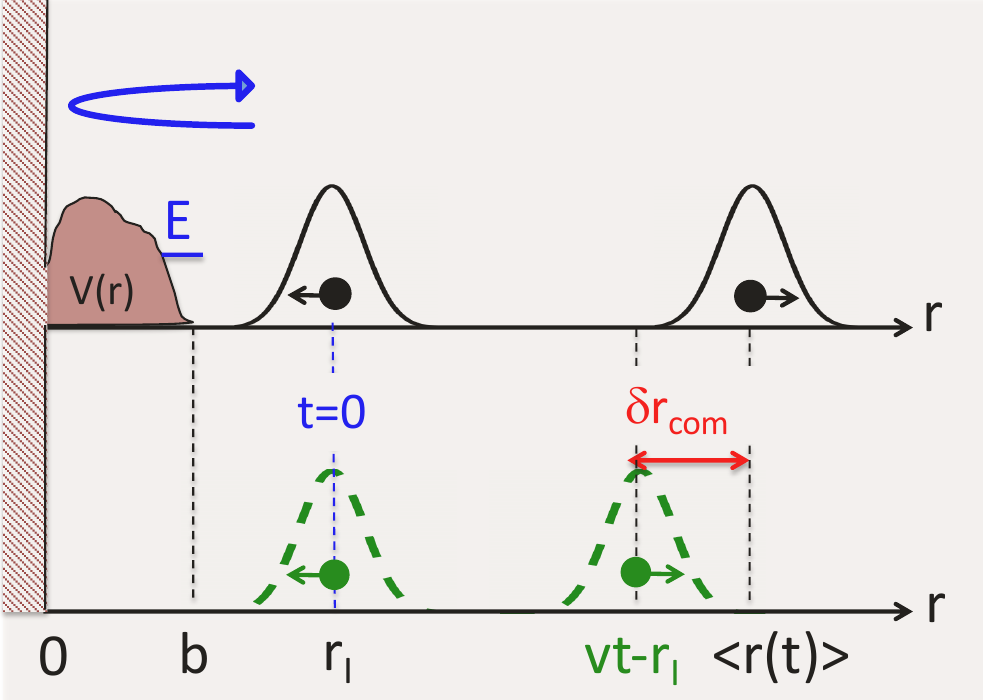}
\centering
\captionof{figure}{A quantum particle with zero angular momentum, $L=0$,  is scattered by a spherically symmetric
potential $V(r)$. The boundary condition at the origin is equivalent to putting an infinite wall at $r=0$.
 }
\label{fig_etaTR_5}
\end{figure}
Proceeding as before, we write
\begin{eqnarray} \label{2f}
\psi(r,t) = e^{ipr-iE_pt}\times \n
\int_{-\infty}^\infty G_0(r-r^\prime,t)\eta(r^\prime,p) dr^\prime,\n
\eta(r^\prime,p) = \frac{e^{-ipr^\prime}}{2 \pi} \int_{-\infty}^\infty S(k,V)e^{ikr^\prime} dk,
\end{eqnarray}
where $S(k,V)$, $|S(k,V)|=1$, is the scattering matrix element.
For a zero-range potential,  obtained in the limit
$V(r) =U\Theta_{[0,b]}$, $b\to 0$, $U\to \infty$, $Ub^2\to const$ \cite{BZP},
$S(k,V)$ is given by
\begin{eqnarray} \label{3f}
S(k,\a)=-\frac{k+i\a^{-1}}{k-i\a^{-1}}=-\left [1+\frac{2i\a^{-1}}{k-i\a^{-1}}\right ],
\end{eqnarray}
where $\a$ is the scattering length \cite{BZP}, positive for a well, $U<0$, and negative for a barrier, $U<0$.
The amplitude distribution $\eta(r^\prime,p)$ becomes
\begin{eqnarray}\label{4f}
{\eta}(r^\prime,p)=
     \begin{cases}
      -\delta(r')+ 2\Theta(-r') |\a|^{-1} \times \\ \exp(-ip r^\prime +  r^\prime/|\a|), \text{if}\q \a < 0\\
      -\delta(r') +2\Theta(r') |\a|^{-1}\times \\ \exp(-ip r^\prime -  r^\prime/|\a|),  \text{if}\q \a >0,\\
     \end{cases}
\end{eqnarray}
In the limit of a broad nearly monochromatic WP
we find
\begin{eqnarray} \label{5f}
\delta r\m {\xrightarrow[ \Delta x \to 0] {}} \R\left [\overline {r'}\right ]=\frac{2\a}{1+k^2\a^2}
\end{eqnarray}
where the real valued \e{complex shift}
\begin{eqnarray} \label{6f}
\overline {r'}= \frac{\int r' \eta(r^\prime,p) dr}{\int \eta(r^\prime,p) dr}= -\frac{\partial \varphi(p,\alpha)}{\partial p}
\end{eqnarray}
equals the Eisenbud-Wigner-Smith time delay, multiplied by the particle's velocity $v$. There is no momentum filtering [cf. Eqs.(\ref{1d}) and (\ref{2d})],
 since all momenta
are perfectly reflected at the origin. As in the previous examples, a zero-range well
($\a >0$) advances the scattered particle, while a zero-range barrier delays it.
\section{Conclusions}
Classically, one can evaluate the duration spent by a particle in a given region of space either
by measuring it directly by means of a clock, or by measuring the spatial shift, the distance between the particle and its
freely moving counterpart. Quantally, both quantities are distributed, and each value is endowed with
a complex valued probability amplitude, rather than with a probability itself.
One faces the usual dilemma. An accurate measurement perturbs the particle, and destroys the studied transition.
An inaccurate one may leave the interference almost intact, but the sought value must remain indeterminate
due to the Uncertainty Principle \cite{FeynL}.
The authors of \cite{BZP} are correct in saying that the Larmor clock measurements are related to the
duration spent in the region [cf. Eq.(\ref{1a})].  Measurements  relying of the transmitted particle's position
determine something else, best described in terms of the virtual shifts imposed on the particle by the potential
[cf. Eq.(\ref{3c})].
Indeed, the range of possible durations shrinks as the region becomes smaller [cf. Eq.(\ref{6a})], while
the range of shifts does not [cf. Eqs.(\ref{2c}), (\ref{5e}) and (\ref{4f})].
What the authors of \cite{BZP} appeared to have failed to notice is that a weakly perturbing Larmor clock can only measure
a \e{complex time} \cite{DSB}, essentially the first moment of an alternating amplitude distribution [cf. Eq.(\ref{3ab})].
The Uncertainty Principle warns against treating
 this quantity, or its parts, as physical time intervals \cite{DSel}.
 The case of a zero-range potential is, however, somewhat special.
 There is a single available duration, $\t=0$ [cf. Eq.(\ref{6a})], and no interference to destroy, no matter
 how accurate the clock is.
 Both quantum and classical particles cannot spend a finite duration in an infinitely small volume.
 \newline
Something different occurs if one uses the centre of mass of a broad wave packet as a \e{pointer}
set up to measure the shift
experienced by the particle in a potential. The measured quantity is, indeed, the said shift, yet the measurement is highly inaccurate,
and there is a range of possible values even in the zero-range limit (\ref{6b}).
The observed displacement of the \e{pointer}, $\delta x\m$ coincides with the real part of a \e{complex shift} [cf. Eq.(\ref{5b})].
One cannot know which of the possible shifts actually occurred, just as one never knows which of the two holes
was chosen by an electron in the double-slit experiment \cite{FeynL}.
\newline
There is a difference in observing a phenomenon, and explaining it.
An experimentalist is perfectly entitled to say \e{I see that with a very narrow high barrier in place, a particle arrives at  the
detector on average $|\delta x\m|/v$ seconds later than would do without the barrier.}
He/she cannot, however, say that this happens because the particle has spent extra time in the barrier region.
Another experimentalist may want to check this claim by using a Larmor clock, and find this time to be zero.
\newline 
{\r In brief, we have used the measurement theory's techniques to explain what makes the two approaches to the tunnelling time problem so different. 
Our analysis can also be applied to other problems, such as reflection of particles, and three dimensional elastic scattering. }
\section*{Acknowledgement}
We thank Grant PID2021-126273NB-I00 funded by MCIN/AEI/ 10.13039/501100011033 and by \e{ERDF A way of making Europe}.  We acknowledge financial support from the Basque Government Grant No. IT1470-22. MP acknowledges support from the Spanish Agencia Estatal de Investigaci\'on, Grant No. PID2019-107609GB-I00.
\appendix
\section{}

Consider, in general, complex valued functions $\eta(x')$ and $G(x')$, and evaluate {an average}
\begin{eqnarray} \label{A1}
\la x\ra \equiv \frac{\int dx x \left |\int dx' G(x-x')\eta(x')\right|^2}{{\int dx  \left|\int dx' G(x-x')\eta(x') \right|^2}}
\end{eqnarray}
in the limit where $G(x)$ is very broad, and can be approximated by
\begin{eqnarray} \label{A2}
G(x-x') \approx G(x)-G'(x)x'
\end{eqnarray}
for all relevant $x'$. To the leading order in $G'(x) = \partial_xG(x)$ on finds
 \begin{eqnarray} \label{A3}
\la x\ra \approx \int x|G(x)|^2dx+\R[\overline {x'}]+2\Im[\overline {x'}]\n
\times\int x \Im \left [G^*(x)G'(x) \right] dx
\end{eqnarray}
where the last term vanishes if $G(x)$ is real.
In particular, if $G(x)$ replaced by $\tilde G_0(x,t)$,
 and  $\eta(x')=\eta_T(x',p)$ is given by Eq.(\ref{4b}),
in the limit $\Delta x \to \infty$  we
have [the spreading of the initial WP can be neglected if the WP is very broad,  $\Delta x = 2/\Delta p\to 0$, $G_0(x,t) \approx G_0(x-vt-x_I))$]
 \begin{eqnarray} \label{A4}
\la x\ra_T \approx vt+x_I+ \R[\overline {x'}_T].
\end{eqnarray}
Subtracting $\la x\ra_0$ given by Eq.(\ref{2b})
yields Eq.(\ref{6b}).
\section{}
Consider, for simplicity, a symmetric  short range potential, $V(x)=V(-x)$, $V(x)=0$ for $|x|>a$.
Prepare an initial wave packet to the left of the potential
\begin{eqnarray} \label{B0}
\la x|\psi(0)\ra=\int_{-\infty}^{\infty} A(k) \exp(ikx)dk
\end{eqnarray}
Define the scattering states $|\phi(k)_L\ra$ where the particle is incident from the left,
\begin{eqnarray} \label{B1}
\la x|\phi_L(k)\ra =
\begin{cases}
(2\pi)^{-1/2} [\exp(ikx)+R(k)\exp(-ikx)]\\
  \text{for}\q x<-a,  \\ (2\pi)^{-1/2} T(k)\exp(ikx)\\
  \text{for}\q x>a.
\end{cases}
\end{eqnarray}
Define the scattering states $|\phi(k)_R\ra$ where the particle is incident from the right,
\begin{eqnarray} \label{B2}
\la x|\phi_R(k)\ra = \la -x|\phi_L(k)\ra
\end{eqnarray}
Expand $|\psi_0\ra$ in the complete orthogonal set $\{|\phi_L(k)\ra, |\phi_R(k)\ra\}$,
\begin{eqnarray} \label{B3}
|\psi_0\ra =\int_0^\infty [\la \phi_L(k)|\phi_0\ra +\la \phi_R(k)|\phi_0\ra]dk.
\end{eqnarray}
Using $T(-k)=T^*(k)$,  $R(-k)=R^*(k)$, $|T(k)|^2+|R(k)|^2=1$,
and recalling that as time progresses,
for the evolved state $|\psi(t)\ra$ we find,
\begin{eqnarray} \label{B4}
\la x|\psi(t)\ra =\int_{-\infty}^\infty T(k)A(k)\exp(ikx)\n
\times \exp(-iE_kt)dk, \q x>a,
\end{eqnarray}
and
\begin{eqnarray} \label{B5}
\la x|\psi(t)\ra =\int_{-\infty}^\infty R(k)A(k)\times \n
\exp(-ikx)\exp(-iE_kt)dk + \n
 \int_{-\infty}^\infty A(k)  \exp(ikx) \times \n
\exp(-iE_kt)dk, \q x<-a.
\end{eqnarray}
The last term in (\ref{B5}) is the freely propagating initial WP. It will eventually leave the $x<-a$ region, and can be neglected.
The Fourier transforms ($k\to x$) (\ref{B4}) and  (\ref{B5}) can be rewritten as Eqs.(\ref{3b}) and  (\ref{1e}), which are valid even if the
initial WP has components with $k<0$
\section{}
Suppose $\psi(x)=\int F(k)\exp(ikx)dk$, and we want to evaluate $\la x\ra\equiv \int x|\psi(x)|^2dx/\int |\psi(x)|^2dx$.
Since $x\psi(x)=-i\int F(k)\partial _k[\exp(ikx)]dk=i\int \exp(ikx)\partial _kF(k) dk$ we find
\begin{eqnarray} \label{C1}
\la x\ra =- \frac{\int \Im [F^*(k)\partial_kF(k)]dk}{\int |F(k)|^2dk}
\end{eqnarray}
For a complete transmission $F(k) =T(k)A(k)\exp(-iE_kt)$ and we obtain
\begin{eqnarray} \label{C2}
\la x\ra_T = x_I +\la v(k)\ra_Tt-\bigg <  \frac{\partial \varphi_T(k,V)}{\partial k} \bigg >_T,
\end{eqnarray}
where
 $\la f(k)\ra_T \equiv \frac{\int f(k) |T(k)|^2|A(k)|^2dk }{\int |T(k)|^2|A(k)|^2dk}$,
 $v(k)=\partial_kE_k$,
 and $T(k,V)=|T(k,V)|\exp(i\varphi_T(k,V))$.
 Furthermore, evaluating $\la v(k)-p/m\ra_T$ for $E_k=k^2/2m$ and a Gaussian WP (\ref{4c})
 yields
\begin{eqnarray} \label{C3}
\la v(k)-p/m\ra =\bigg <  \frac{\partial |T(p,V)|}{\partial p}\bigg >\Delta k^2/2m=\n
\la \Im\left [\overline {x'}_{T}(k)\right]\ra\Delta k^2/2m.
\end{eqnarray}
It is readily seen that in the limit
$\Delta x \to \infty$, $|A(k)|^2\to \delta(p)$ and Eq.(\ref{C2}) agrees with Eq.(\ref{A4})
\newline
For reflection we have $\exp(ikx)\to \exp(-ikx)$, and $T(k,V)\to R(k,V)=|R(k,V)|\exp(i\varphi_R(k,V))$.
A similar calculation yields
\begin{eqnarray} \label{C4}
\la x\ra_R = -x_I -\la v(k)\ra_Rt+\bigg <  \frac{\partial \varphi_R(k,V)}{\partial k} \bigg >_R,
\end{eqnarray}
where $\la f(k)\ra_R \equiv \frac{\int f(k) |R(k)|^2|A(k)|^2dk }{\int |R(k)|^2|A(k)|^2dk}$.
We note that similar but not identical equations were obtained, e.g.,  in \cite{Haug}.


\end{document}